\documentclass[useAMS,usenatbib]{mn2e}
\usepackage{amsmath}
\usepackage{times}
\usepackage{graphicx}
\usepackage{aas_symbols}
\usepackage{graphicx}
\usepackage{epstopdf}
\usepackage{ulem}
\usepackage{color}

\newcommand{\beq}{\begin{equation}}
\newcommand{\eeq}{\end{equation}}
\newcommand{\bea}{\begin{eqnarray}}
\newcommand{\eea}{\end{eqnarray}}
\newcommand{\gae}{\lower 2pt \hbox{$\, \buildrel {\scriptstyle >}\over {\scriptstyle
\sim}\,$}} 
\newcommand{\lae}{\lower 2pt \hbox{$\, \buildrel {\scriptstyle <}\over {\scriptstyle
\sim}\,$}}

\begin{document}

\title[SNRs in the Magellanic Clouds]{Radio SNRs in the Magellanic Clouds as probes of shock microphysics}

\author[Barniol Duran, Whitehead \& Giannios]{Rodolfo Barniol Duran$^{1}$\thanks{Email: rbarniol@purdue.edu (RBD), jewhiteh@purdue.edu (JFW), dgiannio@purdue.edu (DG)}, Joseph F. Whitehead$^{1}$\footnotemark[1], Dimitrios Giannios$^{1}$\footnotemark[1] \\
$^{1}$Department of Physics and Astronomy, Purdue University, 525 Northwestern Avenue, West Lafayette, IN 47907, USA}

\date{Accepted; Received; in original form 22 March 2016}

\pubyear{2016}

\maketitle

\begin{abstract}
A large number of supernova remnants (SNRs) in our Galaxy and galaxies nearby have been resolved in various radio bands. This radio emission is thought to be produced via synchrotron emission from electrons accelerated by the shock that the supernova ejecta drives into the external medium. Here we consider the sample of radio SNRs in the Magellanic Clouds. Given the size and radio flux of a SNR, we seek to constrain the fraction of shocked fluid energy in non-thermal electrons ($\epsilon_e$) and magnetic field ($\epsilon_B$), and find $\epsilon_e \epsilon_B \sim 10^{-3}$. These estimates do not depend on the largely uncertain values of the external density and the age of the SNR.  We develop a Monte Carlo scheme that reproduces the observed distribution of radio fluxes and sizes of the population of radio SNRs in the Magellanic Clouds.  This simple model provides a framework that could potentially be applied to other galaxies with complete radio SNRs samples.
\end{abstract}


\section{Introduction}

Multiwavelength observations in our Galaxy and in other nearby galaxies have discovered and resolved hundreds of supernova remnants (SNRs). Theoretical inferences from these observations are not always straightforward. The distances to SNRs in our Galaxy are uncertain, while samples of other galaxies tend to be incomplete, with many remnants being below the detection threshold. Here we study the sample of SNRs found in the Magellanic Clouds (MCs) in the radio band, which provide us with an almost complete sample at a known distance (\citealp{badenesetal10};  hereafter BMD10). 

Radio SNRs are thought to be produced in the shock that is formed as the supernova (SN) ejecta interacts with the external medium. Particles are accelerated in the expanding shock, magnetic fields are amplified, and particles radiate via the synchrotron mechanism (see, e.g., \citealp{woltjer72, chevalier82a, chevalier82b, chevalier98}, and recently, \citealp{dubnerandgiacani15}). Particle acceleration is thought to proceed via diffusive shock acceleration at SNRs shocks (e.g., \citealp{bell78, blandfordandostriker78}). The efficiency of particle acceleration and magnetic field amplification in SN shocks has been studied extensively (e.g., \citealp{reynoldsandellison92, vinkandlaming03, volketal05, uchiyamaetal07, thompsonetal09}), but it continues to be an active field of research. 

In this Letter, we focus on modeling of radio SNRs in the MCs. Using a simple model of electron acceleration at the SN shock and the corresponding synchrotron emission (Section \ref{Toy_model}), we constrain the fraction of shocked fluid energy in non-thermal electrons ($\epsilon_e$) and magnetic field ($\epsilon_B$) for our chosen sample (Sections \ref{Sample} and \ref{Applications}). We will refer to these fractions as the ``microphysical" parameters. In order to explain both the SNR radio fluxes and their observed sizes in the MCs, we develop a simple Monte Carlo scheme that is able to reproduce these quantities (Section \ref{Monte_Carlo}).  This scheme makes use of the SN rate in the MCs, the observed energy distribution of SN explosions as well as the probability distribution of the densities that surround these SNRs (BMD10, \citealp{maozandbadenes10}). We find both a qualitative and quantitive agreement with the observations in the MCs. 

\section{Model for SNR emission and size } \label{Toy_model}

After the SN explosion, the SN blast wave travels with constant velocity (the ``coasting" phase) until it sweeps enough external material and starts to decelerate. This occurs at a radius and time
\bea
R_{\rm dec} &\approx& 7 \times 10^{18} {\rm cm} \, (E_{\rm 51}/n_0)^{1/3} v_9^{-2/3}, \label{R_dec} \\
t_{\rm dec} &\approx& 220 \, {\rm yr} \, (E_{51}/n_0)^{1/3} v_9^{-5/3},
\label{t_dec}
\eea
where $E$ is the kinetic energy of the SN explosion, $n$ is the number density of the external medium, $v$ is the velocity of the ejecta, and we have used the common notation $Q_x = Q/10^x$ in c.g.s units. Once the blast wave starts decelerating, it follows the Sedov-von Neumann-Taylor (ST) phase (e.g., \citealp{taylor46}). The blast wave radius, $R$, and velocity $v$ in this phase are
\bea
\label{R_ST}
R &\approx& 4.3 \times 10^{19} {\rm cm} \, (E_{\rm 51}/n_0)^{1/5} t_4^{2/5}, \\
v &\approx& 570 \, {\rm km/s} \, (E_{\rm 51}/n_0)^{1/5} t_4^{-3/5},
\label{v_ST}
\eea
where $t$ is the observed time since the explosion in units of $10^4$ yr. Since we are interested in nearby galaxies, 
we assume $z=0$ for the cosmological redshift.

Fermi acceleration predicts that the accelerated particle distribution follows a power-law distribution in momentum with slope $p$. For blast wave velocities of typical SNRs and for $2<p<3$, the bulk of the electron energy is contributed by mildly relativistic particles with Lorentz factor of $\sim 2$ (see \citealp{granotetal06, sironiandgiannios13}). The synchrotron emission for an observed frequency $\nu$, where max($\nu_a$,$\nu_m$) $< \nu < \nu_c$, and $\nu_a$, $\nu_m$ and $\nu_c$ are the synchrotron self-absorption, minimum injection and cooling frequencies, respectively, is \citep{sironiandgiannios13}
\bea 
F_{\nu} \approx (80 \, {\rm mJy}) \bar{\epsilon}_{e,-1} \epsilon_{B,-2}^{\frac{1+p}{4}} E_{\rm 51}^{\frac{11+p}{10}} n_0^{\frac{3+3p}{20}} t_{\rm 4}^{\frac{-3(1+p)}{10}} \nu_{\rm GHz}^{\frac{1-p}{2}} d_{\rm 23.5}^{-2},
\label{F_DN}
\eea
where $\bar{\epsilon}_e \equiv 4 \epsilon_e (p-2)/(p-1)$, $d$ is the luminosity distance and the prefactor is strictly valid only for $p\approx2.4$.

As pointed out in \cite{barniolduranandgiannios15}, if the size (here and throughout ``size" refers to the radius $R$ or diameter $D$ of the remnant) of the radio SNR is known, then one can use equation (\ref{R_ST}) to solve for the unknown external density and then substitute it in equation (\ref{F_DN}). This yields 
\bea 
F_{\nu} \approx (70 \, {\rm mJy}) \bar{\epsilon}_{e,-1} \epsilon_{B,-2}^{\frac{1+p}{4}} E_{\rm 51}^{\frac{5+p}{4}} \Big(\frac{D}{30 \, {\rm pc}}\Big)^{-\frac{3+3p}{4}} \nu_{\rm GHz}^{\frac{1-p}{2}} d_{\rm 23.5}^{-2},
\label{F_DN_2}
\eea
which is analogous to equation (12) in \cite{barniolduranandgiannios15}. 
This equation is independent of both the external density and the time since the explosion, which are the two largely unknown quantities. 

\begin{figure}
\includegraphics[trim=3cm 3cm 6cm 3cm,clip=true,width=7.5cm]{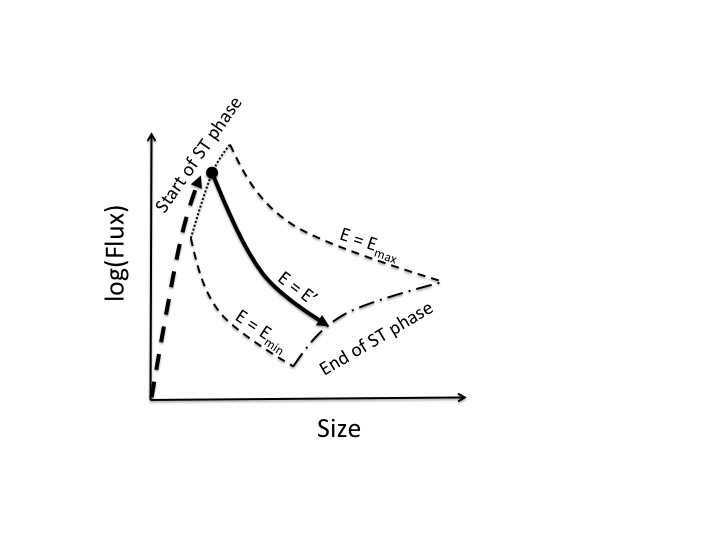}
\caption{Flux-size diagram of a radio SNR. The SNR increases in flux and size during the coasting phase (dashed thick arrow).  For a given external density, the SNR will start decelerating, that is, enter the ST phase somewhere in the thin dotted curve.  The precise location within this curve depends on the SNR energy. As an example, a SNR with energy $E_{\rm min} < E' < E_{\rm max}$ starts the ST phase in the thick dot and will travel along the solid thick arrow as its size increases and its radio flux decreases. A SNR with the largest ($E_{\rm max}$) or smallest ($E_{\rm min}$) energy among SNRs in the same galaxy, will travel along the thin dashed lines, respectively. The SNR follows its respective ST curve until it reaches the radiative phase (end of ST phase, dash-dotted line), which depends on the external density. See  the text in Section \ref{Toy_model} for details.}
\label{fig1}
\end{figure}

In deriving the last expression, we assumed a constant density medium. This assumption is not essential. 
A similar analysis can be carried out in the case of an external medium which follows a power-law density 
in radius with $\propto R^{-k}$. Again, equation (\ref{F_DN_2}) is obtained. Therefore, for given microphysical parameters and SN energy, the SNR flux in the ST phase is
\begin{equation}
F_{\nu} \propto E^{(5+p)/4} D^{-(3+3p)/4} \,\,\,\,\,\,\,\,\,\,\, \text{During the ST phase}. 
\end{equation}
A specific SNR in the ST phase with energy $E$ evolves along a well-defined curve in the flux-size diagram, which follows equation (\ref{F_DN_2}). For a given size, the more energetic the SN, the brighter its emission; conversely, a less energetic SN will have weaker emission, see Fig. \ref{fig1}.

Although equation (\ref{F_DN_2}) directly connects observables to the shock microphysical parameters, it is only valid in the ST phase. As mentioned above, the ST phase commences after the SNR reaches $R_{\rm dec}$, which does depend on external density. Thus, solving for energy in equation (\ref{R_dec}) and substituting in equation (\ref{F_DN_2}) would yield the maximum radio flux for a SNR for a given density $n$, since afterwards the SNR would increase in size but decrease in flux.  This maximum radio flux is 
\begin{equation}
F_{\nu} \propto n^{(5+p)/4} D^3 \,\,\,\,\,\,\,\,\,\,\, \text{Start of ST phase,}
\end{equation} 
as expected, since the flux during the coasting phase increases as $\propto R^3$ as the external medium collected increases in the same manner.  Therefore, a specific SNR that lives in a specific curve of a flux-size diagram starts its ST life in the curve of maximum flux described in this paragraph, see Fig. \ref{fig1}, thus providing a limit on the validity of equation (\ref{F_DN_2}).

The ST phase transitions to a ``radiative" phase, where the blast wave slows down sufficiently that the cooling time of the shock-heated gas becomes less than the age of SNR (e.g., \citealp{blondinetal98}).  The cooling time is given by
\begin{equation}
t_{\rm cool} \approx \frac{3 k_B T}{2 n_{{\rm sh}} \Lambda(T)},
\end{equation}
where $T$ is the temperature behind the shock, $n_{{\rm sh}} = 4 n$ is the density behind the shock, $\Lambda(T)$ is the cooling function at temperature $T$ and $k_B$ is Boltzmann's constant. The temperature behind the shock is 
\begin{equation}
T = \frac{2 (\gamma - 1)}{(\gamma+1)^2} \frac{m_p v^2}{k_B} \approx 3 \times 10^6 {\rm K} \, (E_{51}/n_0)^{2/5} t_4^{-6/5},
\end{equation}
where we used equation (\ref{v_ST}) and $\gamma=5/3$. 
For an $\sim 1/3$ solar metallicity (average value for MCs, e.g., \citealp{russellandbessell1989}) and $T\sim~10^6$ K, the cooling function/temperature dependence can be approximated by $\Lambda \approx 6 \times 10^{-23} \, {\rm erg \, cm^3 \, s^{-1}} \, T_6^{-1}$ (e.g., \citealp{sutherlandanddopita93}).
The time when $t_{cool} \approx t$ yields the ``radiative" time, $t_{\rm rad}$, when the SNR enters the radiative phase and the size of the SNR is $R_{\rm rad}$; these are given by
\bea
\label{t_rad}
t_{\rm rad} &\approx& 2.5 \times 10^4 \, {\rm yr} \, E_{51}^{4/17} n_0^{-9/17}, \\
R_{\rm rad} &\approx& 6.2 \times 10^{19} {\rm cm} \, E_{51}^{5/17} n_0^{-7/17},
\label{R_rad}
\eea
and $R_{\rm rad}$ is the maximum size of a remnant in the ST regime. For a specific energy, SNRs in higher densities turn radiative faster than in lower densities, since the blast decelerates faster.

We can now place another limit on equation (\ref{F_DN_2}), since a SNR will not be able to grow in the ST phase forever.  We can solve for energy in equation (\ref{R_rad}) and substitute it in equation (\ref{F_DN_2}). This yields a limiting flux for the ST phase for a given density $n$, which is 
\begin{equation}
F_{\nu} \propto n^{(35+7p)/20} D^{(35+p)/10} \,\,\,\,\,\,\,\,\,\,\, \text{End of ST phase,}
\label{End_of_ST_phase}
\end{equation}
and this curve marks the end of the ST phase, see Fig. \ref{fig1}. We will assume that the SNR radio flux turns off soon after reaching the radiative stage.  This assumption will be justified for the particular case of the MCs in Section \ref{Applications}.

\section{Sample} \label{Sample}

We consider radio SNRs for which a radio flux and diameter measurement has been obtained. We focus on SNRs with radio non-thermal emission that emit in the optically thin region, that is, their specific flux spectrum is negative.  We are interested in explaining the distribution of radio fluxes and diameters in a specific galaxy. For this purpose, we use the sample in BMD10 (ignoring misidentified objects, \citealp{maggietal16}; and also coasting-phase-SNR 1987A), which includes all radio SNRs detected in the MCs: the Large MC (LMC, at 50 kpc) and the Small MC (SMC, at 60 kpc). The sample consists of 71 radio SNRs. For the MCs, the inferred SN rate is 1 SN every $t_{\rm SN} = 200-500$ yr (e.g., \citealp{maozandbadenes10}). Although SNR sizes are available at different wavelengths, and sizes determined at different wavelengths might disagree (e.g., \citealp{filipovicetal05}), we use the sizes as reported in BMD10. The radio observational limit of the sample lies well below the observed fluxes indicating that most radio SNRs are observed (BMD10). In principle, the technique developed below could be applied to other samples of radio SNRs in other galaxies. 

\section{Constraining the microphysical parameters} \label{Applications}

Using the theory in the previous section, we can attempt to explain the radio SNR data in BMD10. We assume that these SNRs are in the ST phase, and that their flux decreases rapidly right after entering the radiative phase, that is, they essentially disappear. This can be justified the following way (see BMD10 for similar arguments, and also \citealp{fusco84,bandiera10}). A SNR spends approximately $t_{\rm dec} \sim$ 200 yr in the coasting phase, therefore, for a SN occurring every $t_{\rm SN} \sim$ 300 yr, we expect $\lae 1$ SNR to be coasting.  For the ST phase, which lasts for $t_{\rm rad} \sim 3 \times 10^4$ yr, we expect $\sim 100$ SNRs; whereas for the radiative phase, which lasts $\sim 10^6$ yr, we expect $\sim$ 3000 SNRs. We can see that coasting SNRs are too few to account for observations. Radiative SNRs are too many (and therefore must be radio faint). Our expected number of SNRs in the ST phase agrees well with the observed number of SNRs.

\begin{figure}
\includegraphics[width=8.5cm]{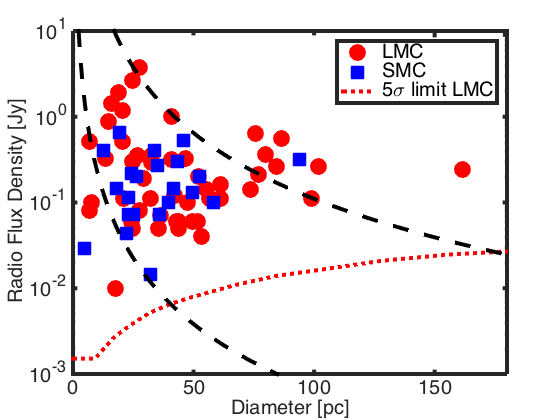}
\caption{Radio flux density of SNRs (most of them at 1.4 GHz) in the MCs versus their size (data from BMD10). The LMC and SMC data are shown in red circles and blue squares, respectively. The SMC fluxes have been scaled to the distance of the LMC.  The red dotted line indicates the 5$\sigma$ radio limit for the selection of candidate SNRs in the LMC (the one for the SMC has been deleted for clarity, although it is a factor of $\sim 3$ larger).  The black dashed curves (bottom-left and upper-right) show the predicted radio fluxes of our model for the observed range of SN energies of $E_{\rm min} = 0.5$ and $E_{\rm max} = 8$ foe, respectively \protect\citep{hamuy03}. Choosing microphysical parameters $\epsilon_{e,-1} = \epsilon_{B,-2} =1$ and $p=2.4$ yields SNRs limiting fluxes that encompass the majority of the data. The largest SNR is DEM L203, recently categorized as an unconfirmed candidate \protect\citep{maggietal16}.}
\label{fig2}
\end{figure}

As discussed above and evident from equation (\ref{F_DN_2}) the radio flux in the optically thin regime and observed frequency $\nu$ that satisfies max($\nu_a$,$\nu_m$) $< \nu < \nu_c$ (satisfied for typical SN parameters) depends only on microphysical parameters and on energy.  As a zeroth order exercise, let us fix the energy range of SNe from observations and that way we can constrain $\epsilon_e \epsilon_B^{(p+1)/4}$. \cite{hamuy03} finds that core-collapse SNe (including both Type II and Type Ib/c) show a distribution of energy from 0.5 to 8 foe (1 foe = $10^{51}$ erg), not including hypernovae, which show even larger energies; Type Ia SN energy is also within these bounds. Using these minimum and maximum values we can calculate the curves of minimum and maximum fluxes with equation (\ref{F_DN_2}), see Fig. \ref{fig1}. We can compare these with the observations in BMD10, where most of the fluxes are at 1.4 GHz and both the LMC and SMC data have been combined to provide a complete picture of the MCs.  We show our results in Fig. \ref{fig2} for $\epsilon_{e,-1} = \epsilon_{B,-2} =1$ and $p=2.4$. It is encouraging that for the given microphysics and the range of observed SN energies, only 15\% of the data points lie outside of our theoretical bounds. This suggests that the spread in fluxes stems from the range of SN energy, while the microphysical parameters: $\epsilon_e \epsilon_B^{(p+1)/4}$, see equation (\ref{F_DN_2}), might be a constant value for all SNRs in consideration. 

\section{Monte Carlo simulations} \label{Monte_Carlo}

To model the entire population of SNRs in a particular galaxy, we develop a Monte Carlo scheme that follows the population of SNRs throughout the galaxy's history. We assume that there is a SN in the galaxy under consideration every $t_{\rm SN}$ (in years). We can move back in time, and the {\it i}th SN took place $t = i \, t_{\rm SN}$ years ago (remnant's age). We consider SNRs older than $\sim 2 t_{\rm SN}$, so that we ignore of the order of $\sim 1$ SNR in the coasting phase. We consider SNRs as old as $10^3 t_{\rm SN}$; although, this number is unimportant as long as it is very large, since for large values of $i$, the remnants are so old that they are in their radio-faint radiative stage.

For every SN, we use a Monte Carlo scheme to randomly choose its external density and its energy from a distribution. For the external density, we assume a probability distribution $P(n) \propto n^{-1}$ as found in BMD10. We allow the minimum and maximum values of density to be $n_{\rm min} \sim 0.1$ to $n_{\rm max} \sim 10$ cm$^{-3}$. The SN energy is chosen to have a maximum probability at $E'=10^{51}$ erg, and we construct a distribution of the form $P(E) = \exp(-0.7(\ln(E/E'))^2)$, so that the probability of obtaining a SN with $>10^{52}$ erg or $<10^{50}$ erg is less than 2\%. 

For each chosen SN energy and density, we calculate the corresponding radiative time, equation (\ref{t_rad}). During the radiative phase the radio flux decreases rapidly, therefore, $t_{\rm rad}$ serves as an upper limit for the age of the remnant (if the remnant age is older than $t_{\rm rad}$ the SNR is assumed not to be detectable in the radio, see Section \ref{Applications}).  Therefore, if $t < t_{\rm rad}$ we record the radio flux and size of the SNR, which can be calculated with equations (\ref{F_DN_2}) (or \ref{F_DN}) and (\ref{R_ST}), respectively. 

\begin{figure}
\includegraphics[width=8.0cm]{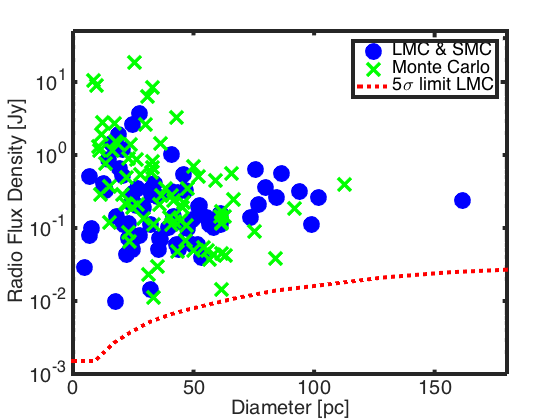}
\caption{Same data as in Fig. \ref{fig2}. Here, the green crosses indicate the result of our Monte Carlo calculation. We have used $\epsilon_{e,-1} = \epsilon_{B,-2} =1$, $p=2.4$, a probability distribution of external density $n$ that is $\propto n^{-1}$ with density from $0.2-10$ cm$^{-3}$, and a probability distribution for SN energy that peaks at $10^{51}$ erg, see Section \ref{Applications}. We have assumed that a SN appears every $t_{\rm} \approx 320$ yr in the MCs (e.g., \citealp{maozandbadenes10}).}
\label{fig3}
\end{figure}

\begin{figure}
\includegraphics[width=8.0cm]{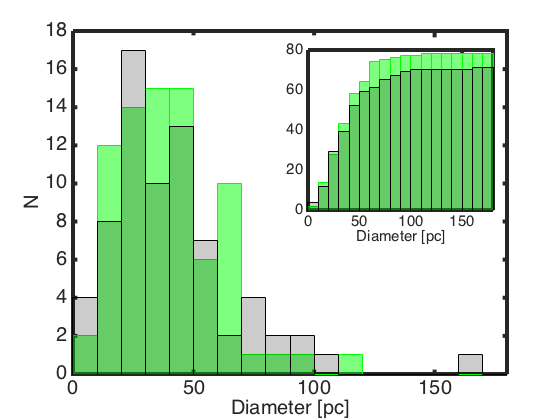} 
\includegraphics[width=8.0cm]{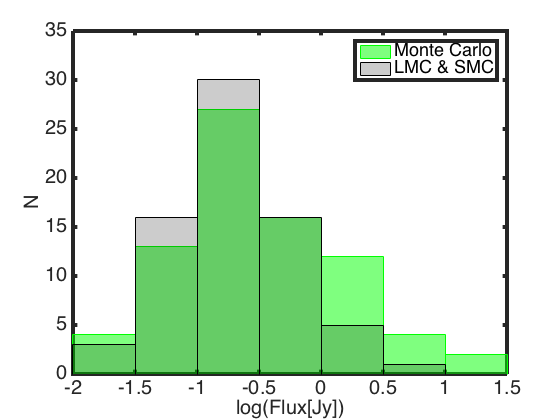} 
\caption{The result of our Monte Carlo calculation (green crosses in Fig. \ref{fig3}) is displayed (light green) along with the data from the LMC and SMC (light gray with black edges) as histograms of sizes (top panel), cumulative sizes (inset) and the radio fluxes (bottom panel), respectively. The darker green regions indicate the overlap between the observed data and the results of our Monte Carlo calculation.}
\label{fig4}
\end{figure}

With this procedure, we can populate the flux-size diagram of the MCs. For all SNRs we fix the microphysical parameters to $\epsilon_{e,-1} = \epsilon_{B,-2} = 1$ and $p=2.4$. We find that a density range from $n_{\rm min} \sim 0.2$ cm$^{-3}$ to $n_{\rm max} \sim 10$ cm$^{-3}$ and $t_{\rm SN} \approx 320$ yr yields good agreement with the data, see Fig. \ref{fig3}. To provide a more comprehensive comparison, we present three histograms in Fig. \ref{fig4}. As can be seen in Figs. \ref{fig3} and \ref{fig4}, our simple model is able to provide a overall explanation for the flux-size distributions in the MCs for fixed microphysics. With the values used, we constrained the microphysical parameters to be $\epsilon_e \epsilon_B^{0.85} \sim 2 \times 10^{-3}$. Below we provide some analytical estimates, which support the choice of parameters.


The results of our Monte Carlo simulations can be understood analytically as follows.  At the end of the ST phase, equation (\ref{End_of_ST_phase}) gives a lower bound to the radio flux. This limit depends on density, so choosing the minimum density $n_{\rm min}$ yields the absolute minimum bound of the ST stage. Since there are only $< 100$ SNRs, this lower limit will be seldom reached with our Monte Carlo calculation (``small" number statistics), so the limit of equation (\ref{End_of_ST_phase}) with $n_{\rm min}$ will be larger by a factor of $\sim 2$. This limit should be smaller than the flux of the largest SNR (see Fig. \ref{fig2}). This constrain yields
\begin{equation}
n_{\rm min} \lae 0.2 \, {\rm cm^{-3}} \epsilon_{e,-1}^{-0.4} \epsilon_{B,-2}^{-0.3} \Big(\frac{F_{\nu}}{0.24 \, {\rm Jy}}\Big)^{0.4} \Big(\frac{D}{160 \, {\rm pc}}\Big)^{-1.4},
\end{equation}
which justifies our density lower limit\footnote{The claim of completeness in BMD10 has recently been put into question (e.g., \citealp{reidetal2015}; \citealp{filipovicandbozzetto2016}). Discovery of fainter SNRs would point to a smaller value of $n_{\rm min}$.}. Since increasing microphysical parameters increases the flux, if we want to maintain the lower bound to the radio flux fixed, then $n_{\rm min}$ should decrease correspondingly.  

In addition, the total number of SNRs in the ST phase can be estimated by knowing the amount of time spent in the ST phase, $t_{\rm rad}$, and the SN rate, as
\begin{equation}
N_{\rm tot} \approx \frac{t_{\rm rad}}{t_{\rm SN}} \approx \frac{2.5 \times 10^4 {\rm yr} \, E_{\rm 51}^{4/17} \bar{n}^{-9/17}}{320 \, {\rm yr}} \sim  70,
\end{equation}
where we used $E_{\rm 51} \sim 1$ and we approximate $\bar{n} = \sqrt{n_{\rm min} n_{\rm max}}$ using $n_{\rm min} \sim 0.2$ cm$^{-3}$ and $n_{\rm max} \sim 10$ cm$^{-3}$. Since a constrain in $n_{\rm min}$ was set above, the previous equation roughly sets a limit on $n_{\rm max}$. We find that $N_{\rm tot}$ agrees well with the results of our Monte Carlo calculation (which is $N_{\rm tot}=78$ SNRs), and also with the observed number of SNRs in the Clouds.

\section{Discussions and Conclusions}

We have studied the population of radio SNRs in the MCs using a simple model to explain their sizes and fluxes. Based on the SN rate in the Clouds, we expect $\sim 100$ SNRs to be in the ST phase, whereas we expect $\sim$ 3000 SNRs in the radiative stage.  Given that the observed sample of radio SNRs are all much brighter than the estimated sensitivity of the radio surveys used, and that the observed number of SNRs is $\lae 100$, it seems that SNRs in the radiative stage are radio faint and that most of the SNRs in the Clouds are in the ST phase (see BMD10). 

In the ST phase, the radio flux of a SNR only depends on its size, microphysical parameters and SN kinetic energy.  It is independent of the external density and the time since the explosion \citep{barniolduranandgiannios15}. Therefore, measuring the SNR size yields a constrain on microphysical parameters that depends solely on SN energy.  The SN energy (of Type Ia, Type II and Type Ib/c SNe), expected to be $10^{51}$ erg, shows a spread that seems to span from 0.5 to 8 foe (e.g., \citealp{hamuy03}). This spread in energy allows for a spread in fluxes, which is roughly consistent with the observed values in the MCs if we fix the microphysical parameters to be $\epsilon_e \epsilon_B^{0.85} \sim 2 \times 10^{-3}$. 

The time of the transition of a SNR to the radiative stage depends on the external density. Since most of the observed SNRs appear to be in the ST phase, the lack of SNRs below a certain radio flux seem to suggest that this is due to their transition to the radiative phase (see BMD10). Knowledge of the external density is necessary to quantify this transition.  For this purpose, we make use of the findings of BMD10 that the probability distribution of external densities in the Clouds varies as $P(n) \propto n^{-1}$. Using this density distribution, the inferred SN rate at the MCs and our flux model, we can explain the distribution of observed fluxes and sizes in the MCs with the same microphysical parameters for all SNRs as mentioned above. This in turn implies that $P(n)$ should extend from $n_{\rm min} \sim 0.1$ to $n_{\rm max} \sim 10$ cm$^{-3}$ in the MCs. We have provided a procedure to populate SNR flux-size diagrams, which might be used to explain the data of other galaxies. Equally important, we have constrained the microphysical parameters for a large number of radio SNR. 

Most of the model parameters are directly constrained by observations. Given the SN rate, the total number of observed SNRs, and realistic probability distributions of external density and SN energy, it appears that the microphysics parameters satisfy $\epsilon_e \epsilon_B^{0.85} \sim 2 \times 10^{-3}$. Although the values of $\epsilon_e$ and $\epsilon_B$ cannot be separately determined with our method, assuming that the shock downstream is in equipartition between non-thermal electrons and magnetic field $\epsilon_e \sim \epsilon_B$, one can infer $\epsilon_e \sim 0.03$, i.e., $\sim 3$\% if the dissipated energy at the shock goes into $\sim 1$~MeV electrons.\footnote{Radio SNe are routinely modeled at the very early phases of their life ($\lae$ months time-scale) within synchrotron emission from the SN blast wave and similar values for microphysics parameters have been found (e.g., \citealp{chevalierandfransson06}).} A smaller fraction $\epsilon_e (m_e/m_p)^{p-2}\sim 10^{-3}$ of the energy is injected into electrons with energy $\sim$ GeV. Assuming that $\sim 10$\% of the energy goes into non-thermal protons, the inferred electron-to-proton injection ratio at $\gae$ GeV energies is $K_{\rm ep}\sim 0.01$. This value is in agreement with the observed cosmic ray composition at Earth (e.g., \citealp{meyer69,picozzaetal13}). On the other hand, if we assume equipartition between the non-thermal protons and magnetic field, $\epsilon_p \sim \epsilon_B \sim 0.2$, then one finds $\epsilon_e \sim 0.008$ and $K_{\rm ep} \sim 0.002$ in agreement with particle-in-cell simulations (e.g., \citealp{parketal15}) and modeling of young SNRs at X-rays and GeV energies (e.g., \citealp{volketal05}).

For a typical SNR size of $D \sim 30$ pc (see Fig. \ref{fig4}), the magnetic field $B^2/8\pi = \epsilon_B 2 n m_p v^2$ is $B \sim 60 \epsilon_{B,-2}^{1/2} \, \mu$G, see equations (\ref{R_ST}) and (\ref{v_ST}). Given that the magnetic field in the MCs is $\sim 1 \, \mu$G (e.g., \citealp{gaensleretal05, maoetal08}), it appears that the amplification needed in MCs SNRs is by a factor of $\gae 10$, although we cannot constrain $\epsilon_B$ uniquely. Interestingly, a similar amplification factor has been found for gamma-ray burst relativistic shocks (e.g., \citealp{barniolduran14, santanaetal14}).

There are hints that the assumption of constant microphysical parameters for all SNRs is over-simplistic. For instance, our simulations produce a few too many bright, young remnants and slightly underproduce the flux of large old ones. This may indicate that particle acceleration tends to be {\it more} efficient for lower remnant speeds. Nevertheless, our simple model provides a powerful diagnostics of particle acceleration and magnetic field amplification in SNRs. 

\section*{Acknowledgements}

We thank Laura Chomiuk, Pawan Kumar, Brian Metzger and Lorenzo Sironi for useful discussions. We acknowledge support from NASA grant NNX16AB32G.



\end{document}